\documentclass[%
 reprint,
nofootinbib,
nobibnotes,
 amsmath,amssymb,
 aps, prd,
 longbibliography,
floatfix,
]{revtex4-2}

\usepackage{graphicx}% Include figure files
\usepackage{dcolumn}% Align table columns on decimal point
\usepackage{bm}% bold math
\usepackage[english]{babel}
\usepackage{fancyhdr}
\usepackage{amsmath}
\usepackage{amssymb}
\usepackage{amsfonts}
\usepackage{psfrag}
\usepackage[applemac]{inputenc}
\usepackage[bf,footnotesize]{caption}
\usepackage[dvipsnames]{xcolor}
\usepackage[colorlinks=True, citecolor=blue, linkcolor=blue, urlcolor=blue,linktocpage]{hyperref}
\usepackage{amsthm}
\usepackage{gensymb}
\usepackage{subfig}
\usepackage{physics}
\usepackage{array}
\usepackage{tcolorbox, mathtools}
\usepackage{soul}
\tcbuselibrary{skins}
\newtcolorbox{mybox}[1][]{before=\centering, drop fuzzy shadow, enhanced, colframe=blue, fonttitle=\bfseries, title=#1, center title}
\usepackage{aas_macros}

\newcommand{\hie}{hierarchical }

\begin{document}

\title{A supplementary radiation-reaction force between two binaries}% Force line breaks with \\

\author{Adrien Kuntz}
 \email{adrien.kuntz@sns.it}
\affiliation{%
 Scuola Normale Superiore, Piazza dei Cavalieri 7, 56126, Pisa, Italy
}%
\affiliation{INFN Sezione di Pisa, Largo Pontecorvo 3, 56127 Pisa, Italy}

\date{\today}% It is always \today, today,

\begin{abstract}
Radiation-reaction forces originating from the emission of gravitational waves (GW) bring binaries to close proximity and are thus responsible for virtually all the mergers that we can observe in GW interferometers. We show that there exists a supplementary radiation-reaction force between two binaries interacting gravitationally, changing in particular the decay rate of the semimajor axis under the emission of GW. This new \textit{binary-binary} force is in some settings of the same order-of-magnitude than the usual 2.5PN force for an isolated binary and presents some striking features such as a dependence on retarded time even in the post-Newtonian regime where all velocities are arbitrarily small. Using Effective Field Theory tools, we provide the expression of the force in generic configurations and show that it interpolates between several intuitive results in different limits.
In particular, our formula generalizes the standard post-Newtonian estimates for radiation-reaction forces in $N$-body systems which are valid only in the limit where the GW wavelength goes to infinity.
%showing that it reduces to the well-known 2.5PN radiation-reaction force for a $N$-body system only when the size of the GW wavelength is greater than the distance between the two binaries.
\end{abstract}

\maketitle

%\tableofcontents

\section{Introduction}
%TODO : cite higher order PN for long term evol of eccentricity of binary?
%TODO : say in abstract system do not lose energy by gravitational radiation in most extreme case ?

The quadrupole formula, derived by Einstein more than one century ago, is probably one of the most useful equations for gravitational-wave (GW) physics since it models the lowest-order GW radiation for any moving mass distribution.
%Among its wide range of applications, one can find CITE CITE. 
In a binary system, the loss of energy by emission of quadrupolar gravitational waves leads to a progressive reduction of its eccentricity and semimajor axis following the Peter-Mathews formula~\cite{PhysRev.131.435}; the force responsible for such a loss of energy has been derived by Burke and Thorne~\cite{1969ApJ...158..997T, 1970PhRvA...2.1501B}. It is thought that all of the binary black hole (BBH) and neutron stars (NS) mergers we see today in LIGO/Virgo interferometers originate from widely separated binaries brought to proximity via quadrupolar emission of gravitational waves during astronomical times~\cite{Maggiore:1900zz}. 

In this respect, it is crucial to correctly account for any modification to binary dynamics, as it could influence the expected merger rate, the parameter distribution (e.g., eccentricity) or the waveform of binaries. Perfectly isolated two-body systems do not exist in Nature, and environmental effects such as the presence of Dark Matter~\cite{Cardoso:2022whc, Eda:2013gg, Macedo:2013qea, Kavanagh:2020cfn, Speeney:2022ryg}, of an accretion disk~\cite{Caputo:2020irr, Toubiana:2020drf, Speri:2022upm, Cardoso:2019rou} or of other celestial bodies in the vicinity~\cite{Antognini:2013lpa, Seto:2013wwa, Stephan_2016, PhysRevD.84.044024, Hoang:2017fvh, Naoz_2013, 2020ApJ...901..125D, Kuntz:2021ohi, Kuntz:2022onu} can affect a binary system. This last possibility in particular can considerably modify the dynamics of a two-body system, e.g. by inducing Kozai-Lidov oscillations~\cite{Kozai:1962zz,1962P&SS....9..719L,Naoz_2016}, resonances~\cite{PhysRevD.105.024017,Bonga:2019ycj, Gupta:2022fbe} or Doppler shifts in the waveform~\cite{Kuntz:2022juv, Tamanini_2020, Robson_2018, 2019MNRAS.488.5665W, Randall_2019, Bonvin_2017, Inayoshi:2017hgw, Strokov:2021mkv, Sberna:2022qbn, 2019PhRvD..99b4025C}. From astronomical observations, many-body systems are very common in the universe~\cite{1997A&AS..124...75T,2014AJ....147...87T, 2008MNRAS.389..925T}, so that they could constitute a non-negligible fraction of the mergers observed in GW observatories~\cite{OLeary:2016ayz, 2020ApJ...903...67M}.

All of the many-body effects mentioned above concern the \textit{conservative} dynamics of $N$-body systems, i.e. the modifications to the binary dynamics that can be computed with a (conserved) Hamiltonian. However, very little is known about the \textit{dissipative} dynamics of many-body systems, i.e. the way in which they emit gravitational waves. This lack of understanding is quite surprising given that it is ultimately this dissipative dynamics which brings the binaries to close proximity and make them merge.
While post-Newtonian potentials are known to 2.5PN in $N$-body systems~\cite{schaeferThreebodyHamiltonianGeneral1987, jaranowskiRadiativePostNewtonianADM1997, koenigsdoerfferBinaryBlackholeDynamics2003, Galaviz:2010te}, it has also been shown that the standard quadrupole formula can give incorrect results in three-body systems if one is not careful about the definition of the center-of-mass~\cite{Bonetti:2017hnb}. This breakdown of the standard post-Newtonian formulaes is linked to the assumption that all bodies lie withing their Near Zone, a sphere of radius of the GW wavelength positioned on the center-of-mass of the system. Indeed, while in a two-body system this assumption is always true, in a many-body system one can easily violate it by placing one body very far from the center-of-mass of the system. In this case, no generic description of the dissipative dynamics in these systems has ever been derived.

\begin{figure}
	\centering

	\includegraphics[width=0.7\columnwidth]{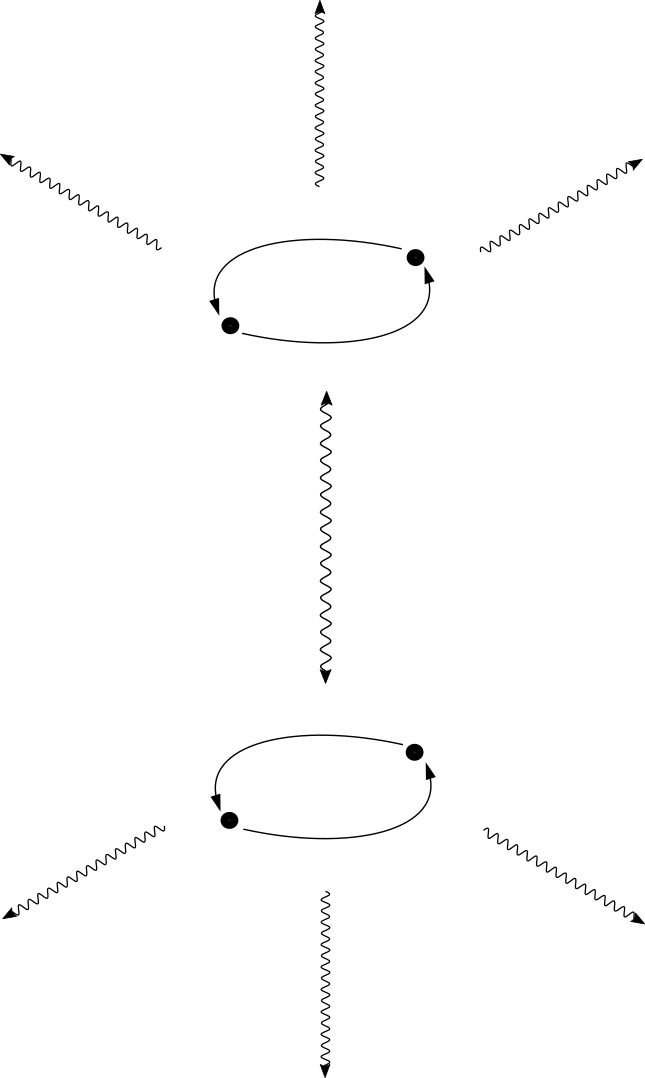}
	
\caption{Illustration of the physical effect which we are going to study in this article. A binary in isolation emits quadrupolar gravitational waves at lowest order; when a second binary system is in the vicinity, a new interaction between the two quadrupoles of the binaries builds up, thus modifying gravitational wave emission.}
\label{fig:illustration}
\end{figure}

In this article, we will focus on a particular type of many-body system where this issue is particularly exacerbated, consisting in two binaries orbiting each other in a \hie configuration (see Figure~\ref{fig:illustration}). In our galaxy, this setting is the most common example of quadruple star systems and their occurrence is not much lower than of triple systems~\cite{2008MNRAS.389..925T}, while double binary systems of BH have already been shown to possibly constitute a relevant fraction of LIGO/Virgo detections~\cite{Liu:2018vzk}. Furthermore, in globular clusters hundreds to thousands of BBH are packed in a zone smaller than $1$Pc~\cite{Rodriguez:2015oxa}, which means that any binary system will be surrounded by a cloud of BBH at relatively short distances.

By using methods borrowed from the Effective Field Theory (EFT) approach to the two-body problem~\cite{goldberger_effective_2006, Goldberger:2007hy, porto_effective_2016, galley_radiation_2009}, we will show that a particular kind of quadrupolar interaction arises between these two binaries. This new force -- which we will call the \textit{binary-binary} radiation-reaction force -- can be associated to a common emission of gravitational waves by the two binaries; it is similar in many aspects to the Burke-Thorne formula which we mentioned before, with some fundamental differences that we will highlight in this article.
One can have a physical intuition on the origin of this new force by looking at Figure~\ref{fig:illustration}: while the standard Burke-Thorne radiation-reaction force emanates from the emission of quadrupolar gravitational waves by an isolated binary, the new effect discussed in this article stems from a binary-binary transfer of gravitons. %We believe that we will give here the first complete expression for this supplementary binary-binary force.

%(in particular it has the same order-of-magnitude, which makes it very relevant for GW astronomy)

% We will show that, beyond the usual well-known Burke-Thorne formula for the quadrupole radiation-reaction force acting on each binary, there exists a new dissipative force applied on both binaries and whose expression will be given in Eq.~\eqref{eq:newRRAction}.
% %whose origin is related to an exchange of radiative gravitons between them. 
% One can have a physical intuition on the origin of this new force by looking at Figure~\ref{fig:illustration}: while the standard Burke-Thorne radiation-reaction force emanates from the emission of quadrupolar gravitational waves by an isolated binary, the new effect discussed in this article stems from a binary-binary transfer of gravitons.

Let us mention here as a preview two of the most striking features of the binary-binary force. First, this new force can be in principle of the same order-of-magnitude than the usual Burke-Thorne effect responsible for the merging of binaries, which makes it very relevant for GW astronomy. In extreme cases the modification is so strong that the whole system do not longer emit the lowest-order quadrupolar waves, as we will show in Section~\ref{sec:lggR}. In practise, though, the binary-binary force probably requires the two binaries to be in resonance in order to be effective as discussed in Section~\ref{sec:long_timescales}.
%In the most extreme case, the binary-binary interaction even completely suppresses the lowest-order quadrupolar radiation from the system as we will show in Section~\ref{sec:lggR}

%A feature that is perhaps surprising is that this new force can be in principle of the same order-of-magnitude than the usual Burke-Thorne effect, although we will see that in practise it probably require the two binaries to be in resonance to be effective. Nevertheless, it presents some characteristics which would never show up in isolated two-body systems. 
The second remarkable characteristic of the binary-binary force is that, \textit{even in the post-Newtonian regime where all velocities are arbitrarily small}, the force features a dependence on retarded time so that its value at a time $t$ cannot be expressed only with values of the system parameters at the same time. The physical origin of this effect is quite trivial, since it is related to wave propagation in flat spacetime contrary to the tail effect already present in isolated two-body systems~\cite{Blanchet:1987wq, Blanchet:1993ec, Blanchet:1992br, Galley:2015kus} where the curvature of spacetime scatters wavepackets. Nonetheless, conventional treatments of post-Newtonian many-body systems as in~\cite{schaeferThreebodyHamiltonianGeneral1987, jaranowskiRadiativePostNewtonianADM1997, koenigsdoerfferBinaryBlackholeDynamics2003} were missing this feature of the binary-binary force since they were restricted to a particular limit where the GW wavelength goes to infinity, which we will discuss in Section~\ref{sec:lggR}. In a generic setting the force inherently mixes different post-Newtonian orders, so that the naive limit $c \rightarrow \infty$ is not under perturbative control.
%Even if this effect quite trivially emerges from wave propagation in flat space-time, the only equivalent 
%This time non-locality is fundamentally different from the tail effect already present in isolated two-body systems~\cite{Blanchet:1987wq, Blanchet:1993ec, Blanchet:1992br, Galley:2015kus}, and can be intuitively understood from the fact that additional length-scales are present in many-body systems as we will discuss in Section~\ref{sec:breakdown}. This is why conventional treatments of post-Newtonian many-body systems as in~\cite{schaeferThreebodyHamiltonianGeneral1987, jaranowskiRadiativePostNewtonianADM1997, koenigsdoerfferBinaryBlackholeDynamics2003} were able to provide the correct expression for the binary-binary force only in a particular limit, which we will discuss in Section~\ref{sec:lggR}; in a generic setting the force inherently mixes different post-Newtonian orders, so that the naive limit $c \rightarrow \infty$ is not under perturbative control.

Let us now present the organization of this article. In Section~\ref{sec:generic_features} we will introduce our notations for the double binary considered, and we will comment on some generic features of the GW emission in $N$-body systems. In Section~\ref{sec:usualRRForce} we will introduce the tools which we will use to derive the effective action ruling out the dissipative dynamics of the system, and apply them to the well-known case of a binary system in order to recover the Burke-Thorne formula. We will then apply this formalism in Section~\ref{sec:new_effective_action} in order to derive the radiation-reaction force stemming from a graviton exchange between the two binaries, which is the main result of this article. The reader interested only in the expression of the binary-binary force can jump directly to Eq.~\eqref{eq:newRRAction} and the subsequent discussion. We will show that we recover standard results in several different limits and finally comment on the effect of the new force on long timescales. %Finally, in Section~\ref{sec:num} we will numerically integrate the equations of motion for a double binary system subject to the new force, and comment on the modifications to the dynamics. 
We will use the $(-+++)$ metric convention, choose units in which $c=1$, and denote Newton's constant by $G$.

%This paper is organized as follows. BLA BLA. The reader in a hurry can go directly to BLA BLA. Convention $G$ and $c$.

\section{Generic features of the dissipative dynamics in a $N$-body system}\label{sec:generic_features}
\subsection{Action and parameters}~\label{sec:params}

We consider a system of four pointlike particles interacting through gravity in General Relativity, representing BH or NS (see Fig~\ref{fig:illustration}). We place ourselves in the physically relevant situation where the objects are organized into two well-separated binaries, which we call the \hie assumption; our choice of indices $N=1\mathrm{A}, 1\mathrm{B}, 2\mathrm{A}, 2\mathrm{B}$ for the point-particles reflects this particular setting. We call the two binary orbits the \textit{inner} orbits, while the much slower motion of the two binaries as they orbit each other is referred to as the \textit{outer} orbit. The spatial positions of the point-particles are denoted by $\bm y_N$ and their spatial velocities by $\bm v_N$; the four positional variables can be split into two center-of-mass positions and two relative coordinates as follows:
\begin{align} \label{eq:def_CM}
    \begin{split}
    \bm Y_{1} &= \frac{m_{1\mathrm{A}} \bm y_{1\mathrm{A}} + m_{1\mathrm{B}} \bm y_{1\mathrm{B}}}{m_{1\mathrm{A}}+m_{1\mathrm{B}}} \; ,\\
    \bm Y_{2} &= \frac{m_{2\mathrm{A}} \bm y_{2\mathrm{A}} + m_{2\mathrm{B}} \bm y_{2\mathrm{B}}}{m_{2\mathrm{A}}+m_{2\mathrm{B}}} \; , \\
    \bm r_1 &= \bm y_{1\mathrm{A}} - \bm y_{1\mathrm{B}} \; , \\
    \bm r_2 &= \bm y_{2\mathrm{A}} - \bm y_{2\mathrm{B}} \; , \\
    \bm R &= \bm Y_{1} - \bm Y_{2} \; .
    \end{split}
\end{align}
The Newtonian definition of the center-of-mass used in~\eqref{eq:def_CM} will be sufficient for our present purposes. Velocities are similarly defined, e.g. $v_1 = \bm v_{1\mathrm{A}} - \bm v_{1\mathrm{B}}$.
We will take the parameters of the two binaries to be similar so that we will use the order-of-magnitude estimates $r_1 \sim r_2 \sim r$, $v_1 \sim v_2 \sim v$ throughout the article. The wavelength of the GW radiation emitted by any binary is thus $\lambda \sim r/v$.
The complete action governing the dynamics of the system is
\begin{align}\label{eq:action}
    S &= \frac{1}{16 \pi G} \int \mathrm{d}^4x \; \sqrt{-g} R \\
    &- \sum_{N} m_N \int \mathrm{d}t \; \sqrt{-g_{\mu \nu} v_N^\mu v_N^\nu} \; ,
\end{align}
where the first term is the usual Einstein-Hilbert action built out of the Ricci scalar $R$ of the metric $g_{\mu \nu}$, and the second is the action for the point-particles labeled with index $N$ and with four-velocity $v_N^\mu = (1, \bm v_N)$, the metric being evaluated at the point-particle positions. From this action, we want to integrate out the gravitational field in order to obtain a Lagrangian depending on the positions of the point-particles only, from which it will be easy to derive the dynamics of the system; this can be done using the path-integral formalism known as Non-Relativistic General Relativity (NRGR)~\cite{porto_effective_2016, goldberger_effective_2006, Goldberger:2007hy, galley_radiation_2009}. While referring the reader to these references for further details, we will just state here that the path-integral for an isolated binary is usually done in two steps, first by integrating out \textit{potential} gravitons on the lengthscale of the binary $r$, and then \textit{radiative} ones on the lengthscale of GW radiation $\lambda \sim r/v$. Using a multipole expansion of the gravitational field around the center-of-mass of any binary, the first step reduces the matter action at lowest nontrivial order to
\begin{align}\label{eq:pp_action_quad}
    \begin{split}
    &- \sum_{N} m_N \int \mathrm{d}t \; \sqrt{-g_{\mu \nu} v_N^\mu v_N^\nu} \\
    &\rightarrow  -  \frac{1}{2} \int \mathrm{d}t \; Q_1^{ij} R_{0i0j}   + (1 \leftrightarrow 2) \; , %-m_1 \int \mathrm{d}t \; \sqrt{-g_{\mu \nu} V_\mathrm{CM,1}^\mu V_\mathrm{CM,1}^\nu}
    \end{split}
\end{align}
where $R_{\mu \nu \rho \sigma}$ is the Riemann tensor evaluated at the center-of-mass of the binary,
$Q_1^{ij}$ is the quadrupole moment of the first binary,
\begin{equation}\label{eq:def_quad}
    Q_{1}^{ij} = \mu_1 \bigg( r_1^i r_1^j - \frac{1}{3} r_1^2 \delta^{ij} \bigg) \;,
\end{equation}
and $m_1 = m_{1\mathrm{A}} + m_{1\mathrm{B}}$, $\mu_1 = m_{1\mathrm{A}} m_{1\mathrm{B}} / m_1$. 
Similar formulaes hold true for $(1 \leftrightarrow 2)$. Note that in Eq.~\eqref{eq:pp_action_quad} we have neglected the monopolar and dipolar couplings of the worldline to the gravitational field. This is because
it turns out that, to obtain the dissipative dynamics of binary systems in which we are ultimately interested, one needs only the coupling of gravitons to a \textit{time-dependent} quantity~\cite{galley_radiation_2009}. Thus, monopolar and dipolar couplings of a binary system to gravity do not contribute to the dissipative dynamics, since they involve respectively to the mass and angular momentum of the binary\footnote{Actually, in our four-body system the angular momentum of any binary could vary because of many-body effects. However, this happens on a very long timescale compared to the time of variation of the quadrupole, so we will neglect their contribution to the dissipative dynamics.}. Note finally that in Eq.~\eqref{eq:pp_action_quad} we have integrated out potential gravitons at the level of each inner binary only; there remains potential modes contributing to the energy of the system on the scale of the outer binary $R$, see the next Section for a more detailed discussion on this point.

%Note that to write Eq.~\eqref{eq:pp_action_quad}, we have neglected a coupling of gravity to the angular momentum of each binary (see CITE for more details), which turns out to be irrelevant for our purposes since 

We split the metric between a flat background and a perturbation, $g_{\mu \nu} = \eta_{\mu \nu} + h_{\mu \nu}$, and expand the action~\eqref{eq:action}-\eqref{eq:pp_action_quad} in powers of velocities and of the gravitational field. %using e.g. the power-counting rules developed in. 
The expansion of the Einstein-Hilbert action to quadratic order will define the propagator mentioned in Section~\ref{sec:Keldysh}, while the quadrupolar couplings~\eqref{eq:pp_action_quad} will define the vertex in Section~\ref{sec:vertex}. However, before moving on to the actual computation of the radiation-reaction force, let us make some comments on the qualitative differences of dissipative effects in a $N$-body system with respect to the case of an isolated binary.

\subsection{Breakdown of the post-Newtonian method} \label{sec:breakdown}

% \begin{figure}
%      \centering
%      \begin{subfigure}[b]{0.3\textwidth}
%          \centering
%          \includegraphics[width=\textwidth]{illustration_2-1.png}
% %         \caption{$y=x$}
%          \label{fig:case1}
%      \end{subfigure}
%      \hfill
%      \begin{subfigure}[b]{0.3\textwidth}
%          \centering
%          \includegraphics[width=\textwidth]{illustration_2-2.png}
% %         \caption{$y=3\sin x$}
%          \label{fig:case2}
%      \end{subfigure}
%      \hfill
%      \begin{subfigure}[b]{0.3\textwidth}
%          \centering
%          \includegraphics[width=\textwidth]{illustration_2-3.png}
% %         \caption{$y=5/x$}
%          \label{fig:case3}
%      \end{subfigure}
%         \caption{Three simple graphs}
%         \label{fig:lambdaVSR}
% \end{figure}

\begin{figure}
    \centering
    \subfloat[]{
        \includegraphics[width=0.5\columnwidth]{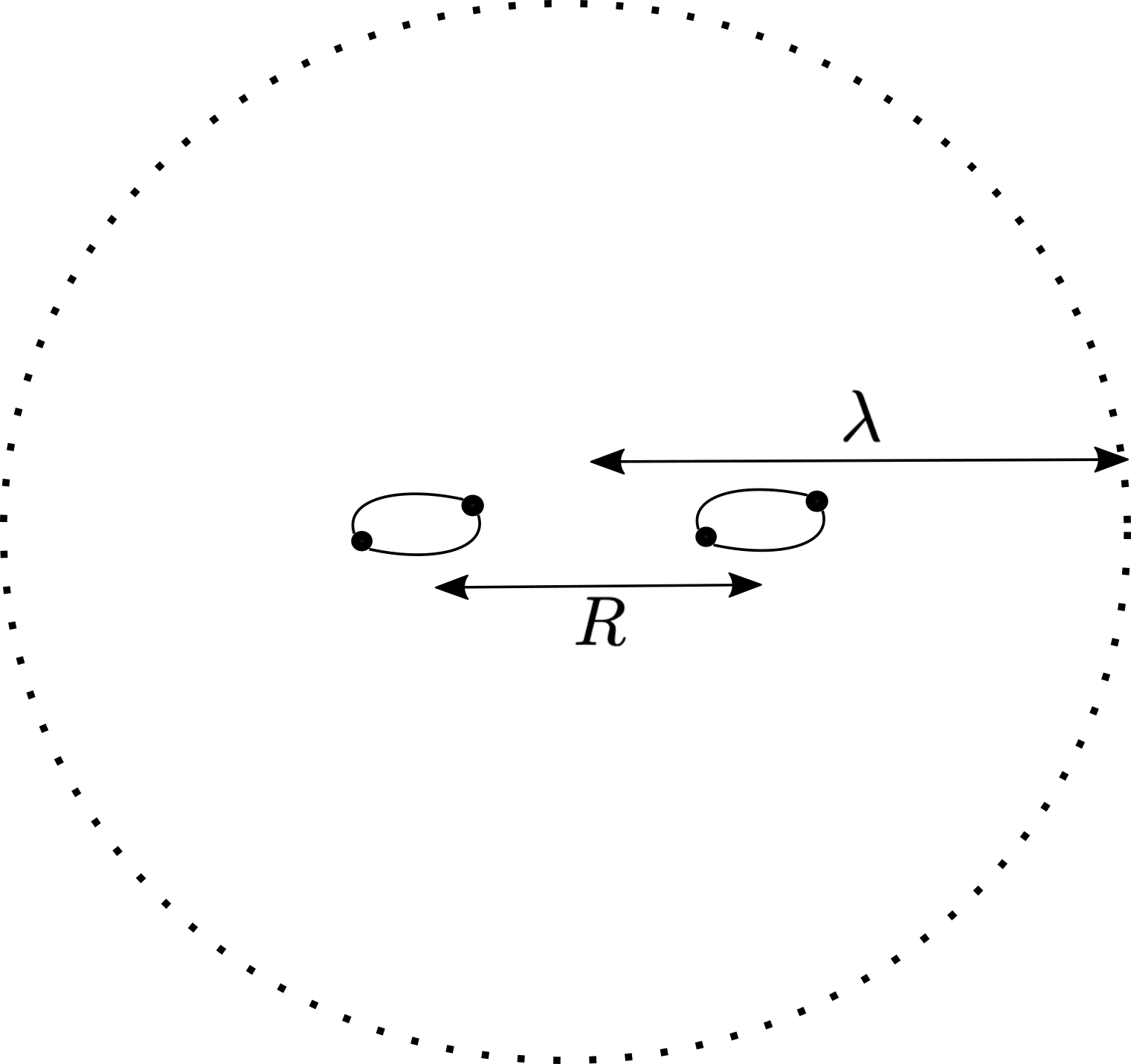}
    } \\
    \subfloat[]{
        \includegraphics[width=0.5\columnwidth]{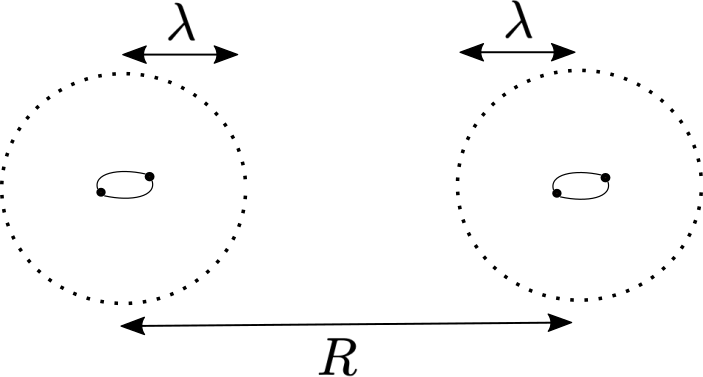}
    }\\
    \subfloat[]{
        \includegraphics[width=0.5\columnwidth]{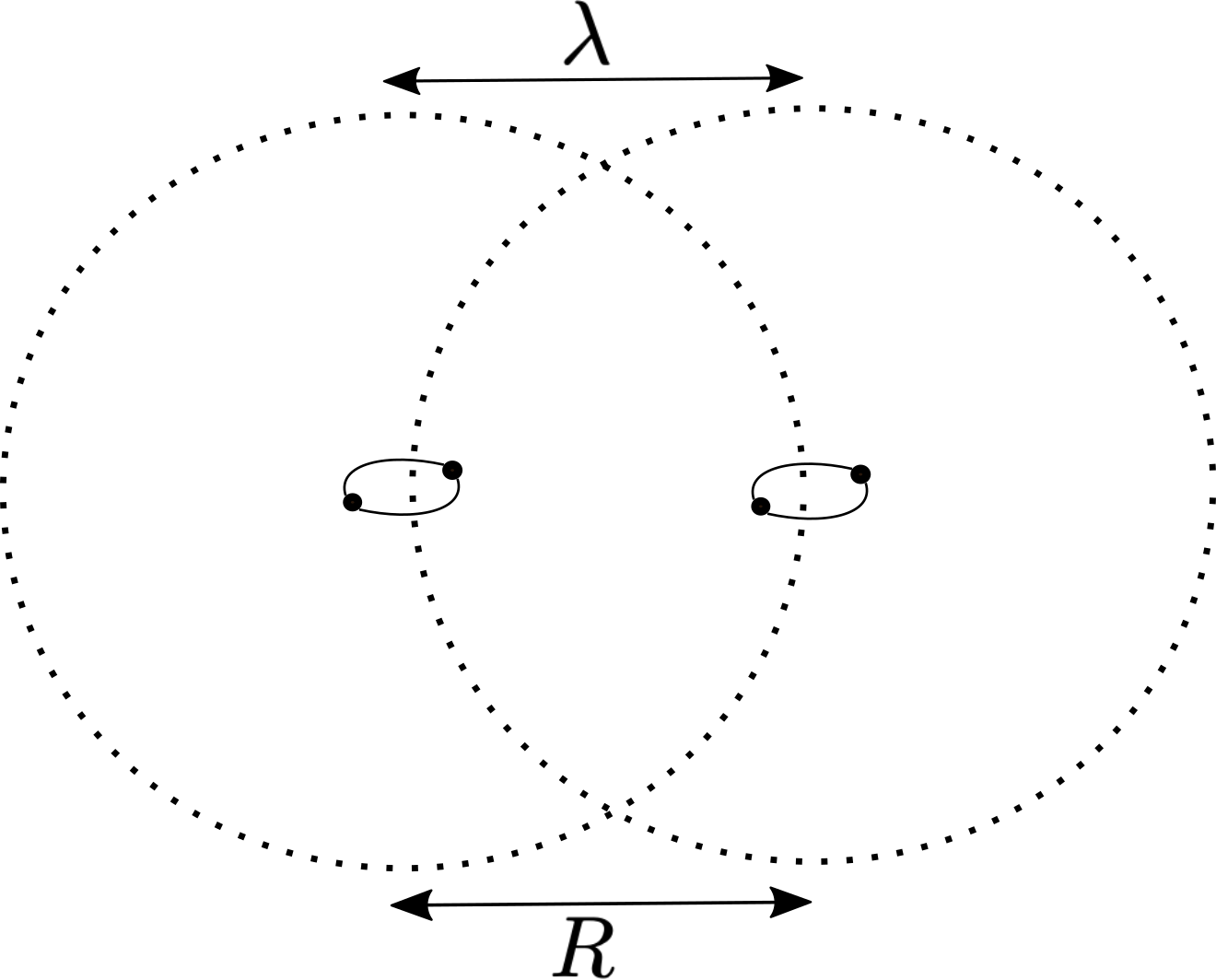}
    }
    \caption{Illustration of the three different limits discussed in Section~\ref{sec:breakdown}. Case (1): the GW wavelength $\lambda$ is bigger than the binary-binary separation $R$. Case (2): $\lambda$ is smaller than $R$. Case (3): $\lambda$ is of the same order-of-magnitude than $R$.}
    \label{fig:lambdaVSR}
\end{figure}

In this Section, we will qualitatively describe the issues that one encounter when one wants to describe GW radiation in a system with $N$ point-like masses when $N>2$ (another, related, viewpoint on the same problem can be found in~\cite{Bonetti:2017hnb}). Let us begin by reviewing the case of a two-body system, and the assumptions at the heart of post-Newtonian expansions which allow to perturbatively solve for the dynamics of the system. The key point is that only to well-separated lengthscales appear in the two-body problem: the typical size of the objects (say $L$) and their relative separation $r$. On top of this, the dynamics depends on only one small parameter $v^2 \sim G m/r$ where $m$ is a typical mass of the objects. This allows one to build a tower of Effective Field Theories (EFT) by integrating out one lengthscale at a time, from the shortest to the longest~\cite{porto_effective_2016}. Thus, beginning with an extended object as a neutron star or a black hole, one "zooms out" from it to represent it as a point particle; one next describes the potential gravitational modes on a lengthscale $r$ reponsible for the conservative dynamics; and finally, one zooms out once more to describe the whole binary system as a point-particle endowed with multipole moments coupled to the radiation field modes varying on a lengthscale $\lambda \sim r/v$, responsible for the loss of energy of the system. We will recall in Section~\ref{sec:usual_effective_action} how this last step can be used to compute the Burke-Thorne radiation-reaction force acting on the binary.

Let us now move on to the kind of four-body systems which we described in Section~\ref{sec:params}. We stress that most of the features which we will describe here are generic to the $N$-body problem as long as $N>2$, and is not particularly tied to our choice of system. Taking the parameters of the two binaries to be similar, one sees from Section~\ref{sec:params} that the system can be described using only one supplementary lengthscale: the separation between the two binaries, which we denote as $R$. One can build several new small parameters out of this lengthscale, e.g. $\varepsilon \sim r/R$ (parametrizing our \hie assumption) or $V^2 \sim G (m_1+m_2)/R$ (parametrizing post-Newtonian effects on the outer orbit). The presence of new small parameters can be used to introduce additional expansions in order to describe the system in an analytic way, such as in~\cite{Kuntz:2021ohi, Kuntz:2022onu}.

One interesting question that one can ask at this point is: are all physical lengthscales of the system well-separated as in the case of a two-body problem, so that one can build a new tower of EFTs similar to the previous one? The answer turns out to be negative because two of these lengthscales do not feature any hierarchy: these are the wavelength of the radiation emitted by any binary $\lambda \sim r/v$, and the separation $R$\footnote{Note that there are also in principle GWs emitted by the outer binary motion, of wavelength $\Lambda \sim R/V$. This scale is much higher than any other lengthscale of the system, and we will send it to infinity in this article (i.e., we do not consider the gravitational waves emitted by the outer binary, which are of smaller amplitude than the GWs of inner binaries)}. At this point one can envision three possibilities (see Figure~\ref{fig:lambdaVSR}):
\begin{enumerate}
    \item $\lambda \gg R$: this means that the Near Zone (as defined in the Introduction) encompasses all objects in the system or, in other words, gravitational waves are emitted in a zone larger than the size of the whole system. In this case, standard post-Newtonian tools can apply, and one can evaluated the energy flux by computing the usual quadrupole formula for a $N$-body system~\cite{jaranowskiRadiativePostNewtonianADM1997,koenigsdoerfferBinaryBlackholeDynamics2003,schaeferThreebodyHamiltonianGeneral1987} using any convenient definition of the center-of-mass of the system to define multipole moments.
    \item $\lambda \ll R$: in this case, one can represent the system by two binaries emitting GWs on a small zone centered around each of them. These GWs then propagate in flat space and are scattered by the other binary. Thus, one can cleany separate the emission process (described by a two-body system in isolation) from the scattering process (scattering of GWs by binaries are described in e.g.~\cite{Annulli_2018, 1979ApJ...233..685T, 1978ApJ...223..285M, Hui:2012yp}). 
    \item $\lambda \sim R$: to our knowledge, there is currently no description of the emission of GWs in this case. Some of the tools used in the two-body problem do not apply any more, while others do. For example, one cannot integrate out first all potential gravitational modes and then radiative ones, since potential modes of the outer binary have a momentum similar to radiation modes of the inner binary. This is responsible for the mixing between conservative and dissipative dynamics which we will describe in Section~\ref{sec:new_effective_action}.
\end{enumerate}

In cases 2) and 3), one should also be careful about retardation effects. Indeed, in the standard post-Newtonian procedure, when deriving the Lagrangian of the system one can systematically treat retardation effects as small perturbations because, for any quantity of interest $f(t)$ evolving on the timescale of the binary $\lambda$, one can write $f(t-r) \simeq f(t) - r \partial_t f + \dots$, and the series converges since $r \partial_t \sim r/\lambda \ll 1$. On the other hand, for $R > \lambda$ one cannot expand $f(t-R)$ as a series at time $t$ and one is obliged to keep any dependence on retarded time. Indeed, the supplementary radiation-reaction force which we will derive in Section~\ref{sec:new_effective_action} will contain a dependence on $t-R$. 

This discussion seems quite abstract for the moment but is in fact very relevant to astrophysics. A $50 M_\odot$ equal-mass BBH able to merge in a Hubble time has an initial semimajor axis of approximately $0.2$AU, which corresponds to a GW wavelength $\lambda \sim 120$AU; while if we also take into account the Kozai-Lidov mechanism in globular clusters, binaries can merge even if their semimajor axis is as high as $10$AU~\cite{Miller:2002pg, VanLandingham:2016ccd}, which corresponds to $\lambda \sim 0.2$Pc. On the other hand, the number density of binary stars in globular clusters can be of the order of $10^4/\mathrm{Pc}^3$ or even more~\cite{grattonWhatGlobularCluster2019, kroupaOriginDistributionBinarystar2001}. For instance, the globular cluster M22 has been suggested to host a population of 5-100 BH in a radius of approximately $1$Pc~\cite{2012Natur.490...71S}, while numerical simulations indicate that hundreds to thousands of BBH can live in these crowded environments~\cite{Rodriguez:2015oxa}.
It thus seems reasonable to expect that in globular clusters or galactic nuclei one or several other BBH or BNS lie within a few radiative wavelength of a binary system, although a detailed study of the double binary population is beyond the scope of this work. %Finally, note that we have observed in our Galaxy an example where two binary star systems orbit each other in such a way that $\lambda/R \sim 10$~\cite{2016A&A...588A.121Z}.

In the rest of this article, we will study a particular type of radiative force in this four-body system and check that our physical intuition is recovered in the limits 1) and 2). We will still use relevant tools pertaining to the two-body problem, and in particular we will model each binary system as a point-particle endowed with multipole moments coupled to the gravitational field as already mentioned in Section~\ref{sec:params}. This is a valid procedure since the size of each binary system is still much smaller than both radiative lengthscales and than the size of the outer binary, so that one can effectively zoom out from each binary. 
This kind of treatment has already been leveraged in~\cite{Kuntz:2021ohi, Kuntz:2022onu} concerning the three body problem, however only in the conservative sector of the dynamics. Finally, note that the new force which we will describe here will only be part of the numerous new many-body interactions in the dissipative sector of a $N$-body problem -- interactions which have not been yet described in full generality in the literature due to this breakdown of the standard post-Newtonian tools mentioned above. However it be seen as a maximal deviation from a two-body problem since we will show that its amplitude is \textit{a priori} of the same size than the standard (Burke-Thorne) radiation-reaction force for an isolated two-body system. A full formalism systematically describing radiation-reaction effects in the 3- and 4-body problem is under preparation by the author.

\section{An EFT derivation of the Burke-Thorne radiation-reaction force}\label{sec:usualRRForce}

%In this Section, we will review how one can compute the dissipative force acting on a binary system originating from the emission of gravitational waves at lowest order. To this aim, we will use NRGR in the Keldysh formalism CITE, which allows one to treat dissipative process in a generalized Lagrangian formalism. In the two-body problem, one can alternatively take a shortcut and avoid the use of Keldysh formalism by computing the loss of energy and angular momentum of the system at infinity and using them in a balance equation. Since ultimately a two-body system is described by two variables (semimajor axis and eccentricity), one can use the equations for the loss of energy and angular momentum to solve for the time-evolution of these two variables. Obviously, two equations are insufficient to solve for the dynamics of a $N$-body system with $N>2$, since these systems are described by many more dynamical variables. This explains why we are obliged to use the full Keldysh formalism in what follows.

In this Section, we will review how one can compute the well-known dissipative force acting on a binary system originating from the emission of gravitational waves at lowest order.
In the case of the two-body problem, the equations governing radiation-reaction decay can easily be found by computing the flow of energy and angular momentum carried by GWs at infinity and using a balance equation. Since ultimately a two-body system is described by two variables (semimajor axis and eccentricity), one can use the equations for the loss of energy and angular momentum to solve for the time-evolution of these two variables. %This is because an isolated two-body orbit can be fully characterized by its energy and angular momentum. 
However, this property does not extend to $N$-body systems with $N>2$. Said differently, the total flow of energy and angular momentum does not provide enough equations to solve for the time-evolution of all the planetary elements of the system. We thus need to use a formalism allowing to compute directly the radiation-reaction force acting on each object, rather than evaluating fluxes at infinity. This is provided e.g. by the Schwinger-Keldysh formalism~\cite{1961JMP.....2..407S, Keldysh:1964ud} (see~\cite{galley_radiation_2009} for a pedagogical introduction in the context of the NRGR formalism); another possibility which we will not follow here is to use the Arnowitt-Deser-Misner (ADM) Hamiltonian formalism of general relativity as in~\cite{jaranowskiRadiativePostNewtonianADM1997,koenigsdoerfferBinaryBlackholeDynamics2003,schaeferThreebodyHamiltonianGeneral1987}.
We will now briefly review the basics constituents of this formalism, while referring the reader to~\cite{galley_radiation_2009} for more details and physical explanations. We will then give the vertex coming from the expansion of the action~\eqref{eq:pp_action_quad} needed to build Feynman diagrams, and we will finally compute the Burke-Thorne radiation-reaction force in this formalism.

\subsection{Setup: the Keldysh formalism} \label{sec:Keldysh}

In the Keldysh or "in-in" formalism, one needs to double each degree of freedom according to $\bm y_N \rightarrow (\bm y_N^{I}, \bm y_N^{II})$ ($\bm y_N$ being the physical coordinates of the $N$-th body) and similarly for the metric perturbation $h_{\mu \nu} \rightarrow (h_{\mu \nu}^{I}, h_{\mu \nu}^{II})$. The action is defined as $S = S^{I} - S^{II}$ ($S^{I}$ depending only on variables $I$)
%and the fields are integrated out using the following matrix of propagators
and the inversion of its part quadratic in perturbations gives rise to the following matrix of propagators
\begin{equation} \label{eq:matrixPropa}
\big\langle h_{a, \mu \nu} h_{b, \alpha \beta} \big\rangle = P_{\mu \nu; \alpha \beta} D_{ab} (x-y) \; ,
\end{equation}
where the indices $a, b = \pm$ are expressed in the Keldysh basis $(+,-)$, obtained by a linear combination of the $(I, II)$ elements:
\begin{equation}\label{eq:changeBasisKeldysh}
f^{+} = \frac{1}{2} \big( f^{I} + f^{II} \big) \; , \quad f^{-} = f^{I} - f^{II} \; ,
\end{equation}
where $f$ is any quantity (possibly vectorial) built out of the particle's trajectories $\bm y_N(t)$ and the metric $h_{\mu \nu}$, entering the vertex in Feynman diagrams. 
Note that the Keldysh indices $a,b = \pm$ are raised and lowered using the 'metric'
\begin{equation}\label{eq:KeldyshMetric}
f_{a}  = c_{ab} f^{b} \; , \quad c_{ab} = \begin{pmatrix}
0 & 1 \\ 1 & 0
\end{pmatrix} = c^{ab} \; ,
\end{equation}
i.e. $f_{\pm} = f^{\mp}$. Indeed, one can note from Eq.~\eqref{eq:changeBasisKeldysh} that $f^{I} g^{I} - f^{II} g^{II} = c_{ab} f^{a} g^{b}$ for any quantities $f$ and $g$ labeled with Keldysh indices.

% \begin{equation}\label{eq:KeldyshMetric}
% f^{a} g_{a}  = c_{ab} f^{a} g^{b} \; , \quad c_{ab} = \begin{pmatrix}
% 0 & 1 \\ 1 & 0
% \end{pmatrix} = c^{ab}
% \end{equation}
% for any quantities $f$ and $g$ labeled with Keldysh indices. Indeed, one can note that $f^{I} g^{I} - f^{II} g^{II} = f^{a} g_{a} $ once the change of basis has been performed, and furthermore that $f^{\pm} = f_{\mp}$.

In equation~\eqref{eq:matrixPropa}, the tensor $P_{\mu \nu; \alpha \beta}$ and the matrix of propagators are defined as
\begin{align}
 P_{\mu \nu; \alpha \beta} &= 16\pi G \big( \eta_{\mu \alpha} \eta_{\nu \beta} + \eta_{\mu \beta} \eta_{\nu \alpha} - \eta_{\mu \nu} \eta_{\alpha \beta} \big) \; , \\
   D_{ab} (x-y) &= \begin{pmatrix}
0 & -i D^{\mathrm{adv}}(x-y) \\ -i D^\mathrm{ret}(x-y) & 0
\end{pmatrix} \; .
\end{align}
Finally, $D^\mathrm{ret}$, $D^\mathrm{adv}$ are the retarded and advanced propagators respectively, given by
\begin{align}\label{eq:propa}
\begin{split}
D^\mathrm{ret}(x-y) &= \int \frac{ \mathrm{d}^4 k}{(2 \pi)^4} \frac{e^{-i k \cdot (x-y)}}{\bm k^2 - (k^0 + i \varepsilon)^2} \\
&= \frac{1}{2\pi} \Theta(x^0 - y^0) \delta \big( (x-y)^2 \big) \; , \\
D^\mathrm{adv}(x-y) &= \int \frac{ \mathrm{d}^4 k}{(2 \pi)^4} \frac{e^{-i k \cdot (x-y)}}{\bm k^2 - (k^0 - i \varepsilon)^2} \\
&= \frac{1}{2\pi} \Theta(y^0 - x^0) \delta \big( (x-y)^2 \big) \; ,
\end{split}
\end{align}
where $(x-y)^2 = \eta_{\mu \nu} (x-y)^\mu (x-y)^\nu$. After integrating out the gravitational field, the effective action will be of the form
\begin{equation}
S_\mathrm{eff} = \int \mathrm{d}t \bigg[ \; \mathcal L \big[ \bm y_N^{I} \big] -  \mathcal L\big[ \bm y_N^{II} \big] + \mathcal L^D \big[  \bm y_N^{I},  \bm y_N^{II} \big] \bigg] \; ,
\end{equation}
where $\mathcal L[\bm y_N]$ is the conservative Lagrangian of the system. %, which has been computed to a high accuracy both for the two-body CITE and the N-body problem CITE. 
Accordingly, any term of the form $\tilde{\mathcal{L}} \big[ \bm y_N^{I} \big] -  \tilde{\mathcal{L}}\big[\bm y_N^{II} \big]$ in the remaining function $\mathcal L^D$ can be absorbed in the conservative dynamics of the system. We will use this property to ensure that $\mathcal L^D$ only contains terms related to dissipative dynamics (this is why we label it with a $D$ subscript). 
%In our setting of a hierarchical three-body system, we will take $\mathcal L$ to be constituted by the 1PN quadrupolar Lagrangian computed in CITE. 
For our purposes, it will be sufficient to take $\mathcal L$ to be the lowest-order Newtonian Lagrangian of the four-body system, so that all post-Newtonian terms will be contained in the dissipative part $\mathcal L^D$.
The equations of motion can be found by varying the effective action with respect to $\bm y_N^{-}$ and taking the physical limit (P.L.) where $\bm y_N^{-} = 0$, $\bm y_N^{+} = \bm y_N$, thus giving
\begin{equation}\label{eq:EOM}
\frac{\mathrm{d}}{\mathrm{d}t} \frac{\partial \mathcal L}{\partial \bm v_N} - \frac{\partial \mathcal L}{\partial \bm y_N} = \left\lbrace \frac{\partial \mathcal L^D}{\partial \bm y_N^{-}} - \frac{\mathrm{d}}{\mathrm{d}t} \frac{\partial \mathcal L^D}{\partial \bm v_N^{-}} \right\rbrace_{\mathrm{P.L.}} \; .
\end{equation}
In this article, we will be only interested by dissipative processes involving the inner orbit of both binaries (stemming from the emission of gravitons at a wavelength $\lambda \sim r/v$); accordingly, there will be no need to label quantities of the outer orbit with Keldysh indices, since their evolution will be dictated by a conservative Lagrangian entering the left-hand side of the equations of motion~\eqref{eq:EOM}. Thus, we will set $\bm Y_{1}^{-} = \bm Y_{2}^- = 0$ and $\bm Y_{1}^{+} = \bm Y_{1}$, $\bm Y_{2}^{+} = \bm Y_{2}$ from the very beginning. 
As an application, one can derive the radiation-reaction force in the case where the dissipative perturbation $\mathcal{L}^D$ is due to the usual two-body quadrupole dissipation,
\begin{align}\label{eq:LBurkeThorne}
\begin{split}
S_\mathrm{BT} &= \int \mathrm{d}t \; \mathcal{L}^D \\
&= - \frac{G}{5} \int \mathrm{d}t \; \bigg[ Q_1^{ij, -} \frac{\mathrm{d}^5}{\mathrm{d}t^5} Q^{+}_{1,ij} + Q_2^{ij, -} \frac{\mathrm{d}^5}{\mathrm{d}t^5} Q^{+}_{2,ij} \bigg] \; ,
\end{split}
\end{align}
%which is nothing but the Burke-Thorne radiation-reaction potential~\cite{1969ApJ...158..997T, PhysRevA.2.1501}, 
where $Q_{1/2}^{ij}$ are the quadrupole moments of binary systems $1$ and $2$, defined in Eq.~\eqref{eq:def_quad}.
This formula has been derived in~\cite{galley_radiation_2009} in the NRGR formalism and we will recover it in Section~\ref{sec:usual_effective_action}. Using now the equations of motion~\eqref{eq:EOM}, we get to the well-known Burke-Thorne radiation-reaction force~\cite{1969ApJ...158..997T, 1970PhRvA...2.1501B} acting e.g. on object $1\mathrm{A}$:
\begin{equation}
    F^i_{1\mathrm{A}} = - \frac{2 \mu_{1} G}{5} \frac{\mathrm{d}^5 Q_1^{ij}}{\mathrm{d}t^5} r_{1, j} \; .
\end{equation}

We will now proceed and show how one can obtain the Burke-Thorne potential~\eqref{eq:LBurkeThorne} in the Keldysh formalism.

\subsection{Vertex} \label{sec:vertex}
%In order to obtain the effective action where gravity has been integrated out along the lines discussed in Section~\ref{sec:params}, one needs the coupling of gravitons to binary systems. It turns out that, to obtain the contribution to the dissipative dynamics in $\mathcal{L}^D$, one needs only the coupling of gravitons to a \textit{time-dependent} quantity~\cite{galley_radiation_2009}. Thus, monopolar and dipolar couplings of gravity to a binary system do not contribute to $\mathcal{L}^D$, since they couple respectively to the mass and angular momentum of the binary\footnote{Actually, in our four-body system the angular momentum of any binary could vary because they can exchange angular momentum. However, this happens on a very long timescale compared to the time of variation of the quadrupole, so we will neglect their contribution to $\mathcal{L}^D$.}.

The quadrupolar coupling to a radiation graviton shown in Eq.~\eqref{eq:pp_action_quad} can be written in the Keldysh formalism as
\begin{equation}\label{eq:quadVertex}
S_{\mathrm{quad}} = - \frac{1}{2}  \int \mathrm{d} t \; Q^{ij, a} R_{i0j0, a} \; .
\end{equation}
In this equation, $Q^{ij}$ can refer to any of the two quadrupole moments $Q_1^{ij}$ or $Q_2^{ij}$ and $R_{\mu \nu \rho \sigma}$ is the Riemann tensor evaluated at the center-of-mass of the corresponding binary.
The vertex~\eqref{eq:quadVertex} will be the only one that we will need in this article.
%Since graviton couplings to conserved quantities like the mass or angular momentum do not radiate~\cite{galley_radiation_2009}, this vertex gives the lowest-order quadrupolar radiation-reaction force from the diagram shown in Fig~\ref{fig:feynDiagr_Jv5}, at order $J v^5$. For completeness, in the next section~\ref{sec:quad} we give the computation of this diagram following the lines of Ref.~\cite{galley_radiation_2009}.

\subsection{Effective action for an isolated binary system}~\label{sec:usual_effective_action}

\begin{figure}
	\centering
	\subfloat[]{
		\includegraphics[width=0.5\columnwidth]{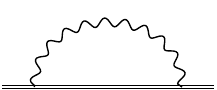}
	}
	
\caption{Feynman diagram for an isolated binary system obtained by integrating out the radiation gravitons at lowest order. The double line represents the inner binary system, the wavy line a graviton, and the vertex can be read from Eq.~\eqref{eq:quadVertex}.}
\label{fig:feynDiagr_isolated}
\end{figure}

We will now compute the lowest-order diagram for an isolated binary system contributing to the dissipative Lagrangian $\mathcal{L}^D$; it stems from the insertion of two quadrupolar vertices shown in Eq.~\eqref{eq:quadVertex}, and it is shown in Figure~\ref{fig:feynDiagr_isolated}. The end result will of course be given by the celebrated quadrupole formula or, more precisely, the Burke-Thorne radiation-reaction potential~\cite{1970PhRvA...2.1501B, 1969ApJ...158..997T} which has been computed in the NRGR formalism in Ref.~\cite{galley_radiation_2009}, and which we have already displayed in Eq.~\eqref{eq:LBurkeThorne}. For simplicity, in this Section we will suppress the indices referring to the first or second binary (e.g. $\bm Y_1 \rightarrow \bm Y)$ since our formulaes will be valid for both of them. Including the symmetry factor of the diagram, one finds using the NRGR rules
\begin{align}
\begin{split}
\mathrm{Fig~\ref{fig:feynDiagr_isolated}} &=  \frac{1}{2} \frac{i^2}{4} \int \mathrm{d}t_1 \mathrm{d}t_2 \; Q^{ij, a}(t_1) Q^{kl, b}(t_2) \\
&\times \big\langle R_{i0j0, a} \big(t_1, \bm Y(t_1) \big) R_{k0l0, b} \big(t_2, \bm Y(t_2) \big)  \big\rangle \; ,
\end{split}
\end{align}
where $Q^{ij}$ is the quadrupole moment of the isolated binary. Note that the Riemann tensor is evaluated at the center-of-mass of the binary $\mathbf Y$.
The derivative structure of the Riemann tensor gives rise to a vertex that we denote as
\begin{widetext}
\begin{align}
\begin{split}
\big\langle R_{i0j0, a} \big(t_1, \bm Y(t_1) \big) R_{k0l0, b} \big(t_2, \bm Y(t_2) \big)  \big\rangle  
= V_{ijkl} \bigg( \frac{\partial}{\partial t_1}, \frac{\partial}{\partial \bm y_A} \bigg) \cdot \left. D_{ab} \big(t_1-t_2, \bm y_A-\bm y_B \big) \right\vert_{\bm y_A =  \bm Y(t_1), \; \bm y_B =  \bm Y(t_2)} \; ,
\end{split}
\end{align}
\end{widetext}
where the definite expression of $V_{ijkl}$ can be found by contracting the gravitons appearing in the Riemann tensor, and is given by
\begin{align}\label{eq:Vijkl}
\begin{split}
    V_{ijkl}(&k^0, \bm k) = 4\pi G \big[ k_i k_j k_k k_l \\
    &- \big(k^0\big)^2 \big(k_i k_k \delta_{jl} + k_i k_l \delta{jk} + k_j k_k \delta_{il} + k_j k_l \delta_{i k} \big) \\
    &+ \big(k^0\big)^4 \big( \delta_{ik}\delta_{jl} + \delta_{il}\delta_{jk} \big) \big] \; ,
\end{split}
\end{align}
where to derive this expression we have used the fact that the quadrupole moment is traceless to remove any term proportional to $\delta_{ij}$ or $\delta_{kl}$. 
To lowest order in the velocity of the center-of-mass $\bm V$, one can set $\bm Y(t_1) = \bm Y(t_2)$. 
Then, performing the Keldysh contraction with the matrix of propagators given in~\eqref{eq:matrixPropa} and using the properties $D^\mathrm{adv}(y-x) = D^\mathrm{ret}(x-y)$ and $V_{ijkl} = V_{klij}$, we are left with
\begin{align}
\begin{split}
\mathrm{Fig~\ref{fig:feynDiagr_isolated} } &= \frac{i}{4} \int \mathrm{d}t_1 \mathrm{d}t_2 Q_{ij}^{-}(t_1) Q_{kl}^{+}(t_2) V^{ijkl} \\
&\cdot\left. D^\mathrm{ret} \big(t_1-t_2, \bm y_A - \bm y_B \big) \right\vert_{\bm y_A = \bm y_B = \bm Y(t_1)} \; .
\end{split}
\end{align}
The integral on $t_2$ can be solved by noting that one can define a master integral whose value within dimensional regularization is (see~\cite{galley_radiation_2009}):
\begin{align} \label{eq:masterInt}
\begin{split}
\int \mathrm{d}t_2 \; &(t_2-t_1)^n \; \mathrm{P.V.} \int \frac{\mathrm{d}^4 k}{(2\pi)^4} \frac{e^{-i k^0(t_2-t_1)}}{\bm k^2 - (k^0)^2} (k^0)^p k_{i_1} \dots k_{i_q} \\
&= \begin{cases} (-1)^n i^{n-1} \frac{n!}{4 \pi (q+1)!!} \delta_{i_1 \dots i_q} & \mathrm{if}\; p+q+1=n \\ 0 & \mathrm{otherwise} \end{cases} \; ,
\end{split}
\end{align}
where $\delta_{i_1 \dots i_q} = \delta_{i_1 i_2} \delta_{i_3 i_4} \dots \delta_{i_{q-1} i_q} + \dots$. Expanding $Q_{kl}^{+}(t_2) = \sum_{n=0}^{\infty} (t_2-t_1)^n (Q_{kl}^{+})^{(n)}/n!$, one finds that only the $n=5$ term contributes and thus
\begin{equation}
\mathrm{Fig~\ref{fig:feynDiagr_isolated} } \equiv i S_\mathrm{BT} = - \frac{i G}{5} \int \mathrm{d}t \; Q_{ij}^- \frac{\mathrm{d}^5}{\mathrm{d}t^5} Q^{ij, +} \; ,
\end{equation}
which recovers the Burke-Thorne formula which we already used in Section~\ref{sec:Keldysh}.

\section{The binary-binary radiation-reaction force}~\label{sec:new_effective_action}
\begin{figure}
	\centering
	\subfloat[]{
		\includegraphics[width=0.5\columnwidth]{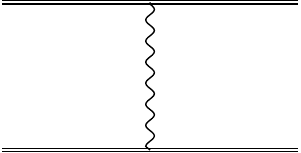}
	}
	
\caption{Feynman diagram giving the binary-binary quadrupolar force. As in the last diagram, the double lines represent the two binary systems and the wavy line a graviton.}
\label{fig:feynDiagr_BB}
\end{figure}
Now that we have introduced our notations and formalism, we will move on to the derivation of the main result of this article. As mentioned in the introduction, in a four-body system one not only has the usual Burke-Thorne dissipative force coming from the self-graviton exchange shown in the Feynman diagram~\ref{fig:feynDiagr_isolated}, but there is on top of this a graviton exchange between the two binaries shown in Figure~\ref{fig:feynDiagr_BB}. This diagram seems \textit{a priori} of same perturbative order as the Burke-Thorne one; however, we will see that it mixes different post-Newtonian orders and furthermore its actual order-of-magnitude value depends on the ratio $R/\lambda$ of the distance between the two binaries and the wavelength of GW radiation. These features emerge because, as stated previously, in a generic setting the momentums of potential gravitons of the outer orbit ($k \sim 1/R$) and radiative gravitons of the inner orbit ($k \sim 1/\lambda$) do not feature any hierarchy, so that to obtain the effective action one is obliged to integrate them out both at the same time. Since radiative and potential gravitons have different post-Newtonian power-counting rules in the NRGR formalism~\cite{goldberger_effective_2006}, this explains the mixing of post-Newtonian orders in the expression for the new force that we will obtain.

\subsection{Derivation of the force}
Let us compute the value of the effective action corresponding to the diagram~\ref{fig:feynDiagr_BB}, which is now quite straightforward to obtain using all the previous ingredients:
\begin{align}
\begin{split}
    i S_\mathrm{BB} &= \frac{i}{4} \int \mathrm{d}t_1 \mathrm{d}t_2 \; Q_1^{ij, -}(t_1) Q_2^{kl, +}(t_2)\\
    &\times V_{ijkl} \bigg( \frac{\partial}{\partial t_1}, \frac{\partial}{\partial \bm Y_{1}} \bigg) \cdot D^\mathrm{ret} \big(t_1-t_2, \bm Y_{1}(t_1) - \bm Y_{2}(t_2) \big) \\
    &+ (1 \leftrightarrow 2) \; ,
    \end{split}
\end{align}
where BB stands for \textit{binary-binary} interaction, which is how we will refer to this new term in the following to distinguish it from the Burke-Thorne one.
Using the expression for the derivative vertex $V_{ijkl}$ given in Eq.~\eqref{eq:Vijkl} and the retarded propgator in~\eqref{eq:propa}, it is easy to evaluate this expression. Note the remarkable feature that the retarded propagator will be evaluated at $\bm R = \bm Y_1 - \bm Y_2$ which is nonzero, so that this diagram is finite and do not need the use of dimensional regularization as was the case for the Burke-Thorne diagram computed in Section~\ref{sec:usual_effective_action}. We obtain

\begin{widetext}
%\begin{mybox}
\begin{align} \label{eq:newRRAction}
    \begin{split}
        S_\mathrm{BB} &= \frac{G}{2} \int \mathrm{d}t \; \frac{Q^{ij,-}_1}{R} \left. \bigg[ \frac{\mathrm{d}^4}{\mathrm{d}t^4} Q_{2,ij}^+ + \frac{2}{R} \frac{\mathrm{d}^3}{\mathrm{d}t^3} Q_{2,ij}^+ + \frac{3}{R^2} \frac{\mathrm{d}^2}{\mathrm{d}t^2} Q_{2,ij}^+ + \frac{3}{R^3} \frac{\mathrm{d}}{\mathrm{d}t} Q_{2,ij}^+ + \frac{3}{R^4} Q_{2,ij}^+ \bigg] \right\vert_{t-R} \\
        &- Q^{ij,-}_1 \frac{R_i R^k}{R^3} \left. \bigg[ 2 \frac{\mathrm{d}^4}{\mathrm{d}t^4} Q_{2,kj}^+ + \frac{8}{R} \frac{\mathrm{d}^3}{\mathrm{d}t^3} Q_{2,kj}^+ + \frac{18}{R^2} \frac{\mathrm{d}^2}{\mathrm{d}t^2} Q_{2,kj}^+ + \frac{30}{R^3} \frac{\mathrm{d}}{\mathrm{d}t} Q_{2,kj}^+ + \frac{30}{R^4} Q_{2,kj}^+ \bigg] \right\vert_{t-R} \\
        &+ Q^{ij,-}_1 \frac{R_i R_j R^k R^l}{2R^5} \left. \bigg[ \frac{\mathrm{d}^4}{\mathrm{d}t^4} Q_{2,kl}^+ + \frac{10}{R} \frac{\mathrm{d}^3}{\mathrm{d}t^3} Q_{2,kl}^+ + \frac{45}{R^2} \frac{\mathrm{d}^2}{\mathrm{d}t^2} Q_{2,kl}^+ + \frac{105}{R^3} \frac{\mathrm{d}}{\mathrm{d}t} Q_{2,kl}^+ + \frac{105}{R^4} Q_{2,kl}^+ \bigg] \right\vert_{t-R} + (1 \leftrightarrow 2) \; .
    \end{split}
\end{align}
%\end{mybox}
\end{widetext}
This is the main result of this article. From this effective action, one can easily derive the radiation-reaction acting on each point-particle force by varying it with respect to coordinates $\bm y_N^-$ present only in the minus component of the quadrupoles $Q^{ij,-}_{1/2}$, and taking the physical limit as in Eq.~\eqref{eq:EOM}. For the time being, the multiple numerical coefficients present in~\eqref{eq:newRRAction} seem completely arbitrary, but we will see in the next Section that they are indeed precisely fine-tuned to yield physically sensible results in the different limits in which we will evaluate this expression. %In the following, we will refer to the new force obtained from the action~\eqref{eq:newRRAction} as the \textit{binary-binary} radiation-reaction force, to distinguish it from the Burke-Thorne one.

Let us comment on several features of the binary-binary force. First, as advertised we see that it indeed contains different post-Newtonian orders: this feature can be seen by e.g. reinstanciating powers of $c$ with dimensional analysis following the rules $\mathrm{d}/\mathrm{d}t \rightarrow \mathrm{d}/c\mathrm{d}t$ and $t-R \rightarrow t-R/c$. As already stated at the beginning of Section~\ref{sec:new_effective_action}, because of the presence of retarded times the perturbative expansion for $c \rightarrow \infty$ is not under control in the generic case, so that one cannot separate the different post-Newtonian orders in Eq.~\eqref{eq:newRRAction} as is usually done in post-Newtonian theory.
Second, we see that this expression is not invariant under time-reversal, which means that it indeed corresponds to a dissipation of energy leaving the system in the form of gravitational waves. However, it is also not odd under time-reversal as was the case for the Burke-Thorne potential~\eqref{eq:LBurkeThorne}, which means that it should contain both a conservative and a dissipative piece, given by the time-symmetric and time-antisymmetric parts of this expression respectively. 
Third, it appears that the order-of-magnitude of the force will ultimately depend on the ratio $R/\lambda$ of the distance between the two binaries and the wavelength of GW radiation, since $\mathrm{d}/\mathrm{d}t \sim 1/\lambda$ (we recall that for order-of-magnitude estimates we take the parameters of the two binaries to be approximately equal). Thus, we will have to split three cases as mentioned in Section~\ref{sec:breakdown} depending on the value of this ratio, which is the subject of the next Section.

\subsection{Physically relevant limits}
In this Section we will consider different limits of the expression~\eqref{eq:newRRAction} and check that the physical intuition developed in Section~\ref{sec:breakdown} is recovered.

\subsubsection{Newtonian limit} \label{sec:newt}
The Newtonian limit can be recovered by reinstanciating powers of $c$ with dimensional analysis and then taking the limit $c \rightarrow \infty$. Replacing time derivatives $\mathrm{d}/\mathrm{d}t \rightarrow \mathrm{d}/c\mathrm{d}t$ and retardation times $t-R \rightarrow t-R/c$, it turns out that the effective action~\eqref{eq:newRRAction} contains a Newtonian (i.e., $\mathcal{O}(c^0)$) piece given by
\begin{align} \label{eq:NewtLimit}
    \begin{split}
        S_{\mathrm{BB}, c \rightarrow \infty} &= \frac{G}{2} \int \mathrm{d}t \; \bigg[ \frac{3}{R^5} Q_1^{ij,-} Q_{2,ij}^+ \\
        &- \frac{30 R_i R^k}{R^7} Q_1^{ij,-} Q_{2,kj}^+ \\
        &+ \frac{105 R_i R_j R^k R^l}{2 R^9} Q_1^{ij,-} Q_{2,kl}^+  \bigg] \\
        &+ (1 \leftrightarrow 2) \; .
    \end{split}
\end{align}
On the other hand, it is quite easy to write the complete effective action of a four-body system in the Newtonian limit: it is given by
\begin{equation}\label{eq:SNewt}
    S_\mathrm{Newt} = \frac{G}{2} \int \mathrm{d}t \; \sum_{N,M} \frac{m_N m_M}{|\bm y_N - \bm y_M |} \; .
\end{equation}

To recover Eq.~\eqref{eq:NewtLimit} from Eq.~\eqref{eq:SNewt}, one has to perform a multipolar expansion of Eq.~\eqref{eq:SNewt} up to quadrupole order in the center-of-mass frame of each binary (as explained in Section~\ref{sec:params}) and keep terms involving interactions betweens the two quadrupole moments which gives exactly Eq.~\eqref{eq:NewtLimit}. We thus see that the Newtonian limit of our binary-binary action is correct. Furthermore, in this limit the action is obviously time-symmetric which means there is no energy loss from the system as there is no gravitational wave emission.

What is the order-of-magnitude of the force in this limit? Of course, the Burke-Thorne radiation-reaction force is parametrically smaller than the binary-binary force for $c \rightarrow \infty$, since it is a purely relativistic effect of order $\mathcal{O}(c^{-5})$. On the other hand, let us compare the amplitude of the binary-binary force with respect to the Kozai-Lidov interaction $S_\mathrm{KL}$ also present in the expansion of the Newtonian action~\eqref{eq:SNewt} in the center-of-mass frame of the binaries (see e.g.~\cite{2013MNRAS.431.2155N, Kuntz:2022onu} for definite expressions), and responsible for many interesting astrophysical phenomena~\cite{Hoang:2017fvh,Wen:2002km,Chandramouli:2021kts,Naoz_2013,Naoz_2016,VanLandingham:2016ccd, Martin_2015,10.1093/mnrasl/slv139,Mazeh_1997, 2014ApJ...785..116L,2015MNRAS.451.1341L,Liu_2014,1979A&A....77..145M}: 
\begin{equation}
    \frac{S_\mathrm{BB}}{S_\mathrm{KL}} \sim \bigg( \frac{r}{R} \bigg)^2 \; .
\end{equation}
Thus, the binary-binary force is actually a subdominant many-body conservative interaction in the Newtonian limit, and can affect the system only on timescales much longer than the Kozai-Lidov time. Still, it can lead to interesting dynamical evolution for the system, as shown in e.g.~\cite{2020MNRAS.493.5583T, 2018MNRAS.475.5215B}.
%and we do not expect it to have a dramatic effect on the dynamics of real astrophysical bodies. 
As a side remark, note that while most Newtonian interactions between the two binaries can be modelled by replacing one binary with a point-particle, the binary-binary force is the lowest-order effect where this approximation breaks down and one has to take into account the finite-size of both binaries, as shown in~\cite{Hamers:2020gbk}.

\subsubsection{$\lambda \gg R$} \label{sec:lggR}
This corresponds to case 1) in Section~\ref{sec:breakdown}: the Near Zone is much bigger than the size of the four-body system, so that one should recover the standard post-Newtonian formulaes for the radiation-reaction forces, see e.g.~\cite{jaranowskiRadiativePostNewtonianADM1997,koenigsdoerfferBinaryBlackholeDynamics2003,schaeferThreebodyHamiltonianGeneral1987}. In the small-$R$ limit, one can expand all retarded times $t-R$ in powers of $R$, schematically $Q_2^{ij}(t-R) = Q_2^{ij}(t) - R \partial_t Q_2^{ij} + \dots$. This series expansion is indeed valid as long as $R \partial_t \sim R/\lambda \ll 1$, and produces as a result a lowest-order radiation-reaction force which only depends on the current time $t$ and not of the past history of the system. %as in the standard post-Newtonian formalism. 

Plugging this expansion in the binary-binary action~\eqref{eq:newRRAction}, we obtain a series of terms proportional to time-derivatives of $Q_2^{ij}(t)$, some of which are even under time-reversal (i.e. they conserve energy) while others are odd (i.e. they dissipate energy). The conservative terms are not so much interesting since, as in the previous Section~\ref{sec:newt}, they are smaller in magnitude than the Kozai-Lidov terms which are the dominant conservative contribution to the dynamics of the binaries. On the other hand, the terms dissipating energy can sensibly alter the dynamics of the system (for example, they can induce a nontrivial evolution of the semimajor axis of the binaries, while conservative terms will keep it constant over long timescales when the system is far from resonances~\cite{Kuntz:2021ohi, Kuntz:2022onu, 2013MNRAS.431.2155N}). Thus we will only keep the lowest-order dissipative term when looking at the small-$R$ expansion of the action while throwing out conservative contributions. As a result, the expanded action is
\begin{equation}\label{eq:smallRLimit}
    S_{\mathrm{BB}, \lambda \gg R} = - \frac{G}{5} \int \mathrm{d}t \; Q_{1,ij}^- \frac{\mathrm{d}^5}{\mathrm{d}t^5} Q_2^{ij, +} + (1 \leftrightarrow 2) \; .
\end{equation}

This formula is quite similar to the Burke-Thorne action~\eqref{eq:LBurkeThorne} (in particular it is of the same $2.5$PN order), and we will comment on this later on. 
Note the remarkable feature that the numerical coefficients in Eq.~\eqref{eq:newRRAction} all conspire to give a lowest-order dissipative force proportional to \textit{five} time-derivatives of $Q_2$, i.e of 2.5PN order as should be the case in the standard post-Newtonian formalism. In fact, one can check that the expansion of the second line of Eq.~\eqref{eq:newRRAction} gives a lowest-order dissipative term of 3.5PN order, while the expansion of the third line gives a dissipative term of 4.5PN order.

How should we interpret the result of Eq.~\eqref{eq:smallRLimit}? Taking also into account the Burke-Thorne action~\eqref{eq:LBurkeThorne}, one gets the total dissipative effective action
\begin{equation}\label{eq:totActionSmallR}
    S_\mathrm{eff} =  - \frac{G}{5} \int \mathrm{d}t \; Q_{1,ij}^- \frac{\mathrm{d}^5}{\mathrm{d}t^5} \big[ Q_1^{ij, +} + Q_2^{ij, +} \big] + (1 \leftrightarrow 2) \; .
\end{equation}
The meaning of this equation is quite clear: the radiation-reaction force is induced by the (fifth time-derivative of the) \textit{total} quadrupole moment of the four-body system, which in the \hie approximation at lowest order just splits in the sum of the quadrupole moments of the two binaries. This is indeed the result that one should expect from already known computations in the post-Newtonian formalism~\cite{jaranowskiRadiativePostNewtonianADM1997,koenigsdoerfferBinaryBlackholeDynamics2003,schaeferThreebodyHamiltonianGeneral1987}. 
The previous expression~\eqref{eq:totActionSmallR}, though, bears some striking consequences. Let us imagine for a moment that the two binaries share the exact same orbital elements, apart from a dephasing $\phi$ in their mean anomaly (see also Appendix~\ref{app} for a generalization to a case where not all planetary elements of the two binaries are equal). It is quite straightforward to obtain an evolution equation for the planetary elements from the Lagrangian using standard techniques, see e.g.~\cite{Kuntz:2021ohi}, and we obtain for the evolution of the semimajor axis of the first binary $a_1$:
\begin{equation} \label{eq:da1dt}
    \frac{\mathrm{d}a_1}{\mathrm{d}t} = \left.  \frac{\mathrm{d}a_1}{\mathrm{d}t} \right\vert_\mathrm{Peters} \big(1+ \cos \big( 2 \phi \big) \big) \; ,
\end{equation}
where $\left. \frac{\mathrm{d}a_1}{\mathrm{d}t} \right\vert_\mathrm{Peters}$ is the standard Peters-Mathew formula for the GW decay of the semimajor axis~\cite{PhysRev.131.435}. Thus, this decay is modulated by the angle $\phi$, and $a_1$ can even stay constant if one chooses $\phi = \pi/2$: such a system completely stops losing energy by gravitational radiation! Indeed, one should remember that we are talking about gravitational \textit{waves}: the wave amplitude at large distance from the four-body system can be found by summing the waves produced by the two quadrupoles; if the dephasing between these two waves is equal to $\pi$ (corresponding to $\phi = \pi/2$ due to the quadrupolar nature of the waves), they can actually interfere destructively to give a zero quadrupolar amplitude. The radiation emitted by the binary would thus be given by the next-to-leading order (octupole and current quadrupole), which would completely change the characteristics of gravitational wave emission. A more detailed analysis of such kind of synchronization was performed recently in~\cite{Seto:2018xnu, Seto:2020lez}, where it was shown that mass transfer by the
Roche lobe overflow of binary white dwarfs could indeed capture the whole system in such a resonant state.
%Even if this choice of parameters seems to be a fine-tuned situation, it has recently been shown that double binary systems can be captured in such a synchronized state
%Of course, this choice of parameters seems to be a very fine-tuned situation, but it would be interesting to know if some dynamical mechanism could bring binaries in such a state (see e.g.~\cite{2022A&A...667A..53P} for an example of a double binary star system in a 3:2 resonance).
%Finally, notice that this modulation of the semimajor axis decay is only possible in this limit $\lambda \gg R$; we will see in the next Section that the binary-binary force is subdominant in the other regime $\lambda \ll R$. %TODO rephrase this

\subsubsection{$\lambda \ll R$}
The large-$R$ limit of Eq.~\eqref{eq:newRRAction} is straightforward to obtain, and we get
\begin{align}\label{eq:largeRLimit}
\begin{split}
    S_{\mathrm{BB}, \lambda \ll R} &= \frac{G}{2} \int \mathrm{d}t \; \frac{Q^{ij,-}_1}{R} \bigg[ \frac{\mathrm{d}^4}{\mathrm{d}t^4} Q_{2,ij}^+ \\
    &- 2 \frac{R_i R^k}{R^2} \frac{\mathrm{d}^4}{\mathrm{d}t^4} Q_{2,kj}^+ + \frac{R_i R_j R^k R^l}{2R^4} \frac{\mathrm{d}^4}{\mathrm{d}t^4} Q_{2,kl}^+  \bigg] \bigg\vert_{t-R} \\
    &+ (1 \leftrightarrow 2) \; .
\end{split}
\end{align}
In this case, one can picture the system as two isolated binaries emitting GWs on a small zone centered around each of them. Each binary emits GW which propagate and then scatter on the other binary. Intuitively, the amplitude of the binary-binary force should be quite weak in this setting. Let us see how we can recover Eq.~\eqref{eq:largeRLimit} using this intuitive reasoning. The quadrupolar waveform at a distance $\bm d$ of an isolated binary is in the TT gauge~\cite{Maggiore:1900zz}
\begin{equation}
    h_{ij}^\mathrm{TT} = \frac{2G}{d} \ddot Q_{ij}^\mathrm{TT}(t-d) \; ,
\end{equation}
where $Q^{ij}$ is the quadrupole moment of the binary and the TT operator project spatial indices in the TT gauge as
\begin{equation}\label{eq:QijTT}
    Q_{ij}^\mathrm{TT} = Q_{ij} - n_j n^k Q_{ik}  - n_i n^k Q_{jk} + \frac{1}{2}n_i n_j n^k n^l Q_{kl} \; ,
\end{equation}
% \begin{equation}
%     \Lambda_{ij;kl} = \delta_{ik} \delta_{jl} - \frac{1}{2} \delta_{ij} \delta_{kl} - n_j n_l \delta_{ik} - n_i n_k \delta_{jl} + \frac{1}{2}n_k n_l \delta_{ij} + \frac{1}{2} n_i n_j \delta_{kl} + \frac{1}{2}n_i n_j n_k n_l
% \end{equation}
where $n_i = d_i/d$. Let us plug this formula with $Q_{ij} = Q_{2,ij}$ in the action of the first binary~\eqref{eq:quadVertex}, treating $h_{ij}$ as an external gravitational field. In this case, we should take the physical limit $h^-_{ij}=0$, $h^+_{ij} = h_{ij}$ in Eq.~\eqref{eq:quadVertex} and set $\bm d = \bm R$ in Eq.~\eqref{eq:QijTT}. Using $R_{0i0j}^\mathrm{TT} = - \ddot h_{ij}^\mathrm{TT} /2$, we see that we recover exactly the term proportional to $Q^-_{1,ij}$ of Eq.~\eqref{eq:largeRLimit}, while the second one is obtained by considering the symmetric situation $(1 \leftrightarrow 2)$.

In the limit that we are considering, the order-of-magnitude of the binary-binary action with respect to the Burke-Thorne one $S_\mathrm{BT}$ is
\begin{equation}
    \frac{S_{\mathrm{BB}, \lambda \ll R}}{S_\mathrm{BT}} \sim \frac{\lambda}{R} \ll 1 \; ,
\end{equation}
which means that the binary-binary force is always subdominant in this regime, as expected. Note that it however still contains dissipation of energy, since due to the presence of retarded time the action~\eqref{eq:largeRLimit} is not time-reversal symmetric.

\subsubsection{$\lambda \sim R$}
 In this case, we can not simplify any further the binary-binary force~\eqref{eq:newRRAction}, as we expect all of its terms to contribute equally. Furthermore, the order-of-magnitude of the binary-binary force is the same as the Burke-Thorne one, so that it can in principle give important contributions to the dissipative dynamics. However, to evaluate the impact of the binary-binary force in this situation one has to solve a complicated delay-differential equation which we leave to future work. We will have more to say concerning the average force in the next Section.

\subsection{Effects on long timescales}\label{sec:long_timescales}
It is a common practise to average either the radiation-reaction or the many-body forces over the orbital timescale of the inner binary in order to obtain simplified equations. This is technically done by parametrizing the coordinates $\bm y_N$ in terms of the osculating orbital elements of the elliptic orbits~\cite{1991ercm.book.....B}: only one of these elements (the mean anomaly) then evolves on the timescale of the binary, while other stay approximately constant during this time due to the \hie assumption. This allows for a straightforward evaluation of the time averages, since one only needs to integrate out the mean anomaly keeping other orbital elements fixed~\cite{Kuntz:2021ohi, Kuntz:2022onu}.
Doing so at the level of the Burke-Thorne force yields the Peter-Mathews equations~\cite{PhysRev.131.435}, while the Kozai-Lidov Hamiltonian is usually obtained by also averaging over the outer orbital timescale~\cite{1962P&SS....9..719L, Kozai:1962zz, 2013MNRAS.431.2155N, Naoz_2016}. In our case, we cannot straightforwardly perform this average directly at the level of the effective action~\eqref{eq:newRRAction}, because it still contains the doubled Keldysh variables $\bm y_N^\pm$ while only the physical coordinates $\bm y_N$ are parametrized in terms of osculating elements. However, it is definitely possible to perform the average on the force once it is derived as in Eq.~\eqref{eq:EOM}. 

Let us now make some general remarks concerning the averaging procedure. The quadrupole moment of each binary (say $Q_1^{ij}$ for definiteness) depend on time through the radius $\bm r_{1}$, which in turn can be expanded for small eccentricities in a Fourier series containing terms like $\cos (p n_1 t + \phi_1)$ where $p$ is an integer, $n_1$ is the frequency of the first binary and $\phi_1$ an arbitrary phase. Since the binary-binary force acting on the first binary consists in a sum of products of quantities depending only on one binary (see Eq.~\eqref{eq:newRRAction}), its time-dependence will be schematically contained in terms like
\begin{equation}
    \cos \big( p n_1 t + \phi_1\big) \cos \big( q n_2 [t-R(t)] + \phi_2 \big) \; ,
\end{equation}
 where $q$ is an integer, $n_2$ is the frequency of the second binary and $\phi_2$ another arbitrary phase. In the \hie assumption we can average this term over time by assuming that any parameter is constant, including the outer orbital radius $R(t)$ which varies on timescales much longer than the inner binaries timescales. We thus obtain that the average vanishes apart when one satisfies the resonance condition $p n_1 = q n_2$. Two cases can be separated:
 \begin{itemize}
     \item Binaries are far from resonance and only the trivial case $p=q=0$ can give a nonzero time-average. This corresponds to performing the average on each binary separately. However, when independently averaging terms depending on $Q_2^{ij}$ in the binary-binary force derived from Eq.~\eqref{eq:newRRAction}, one finds that all time-derivatives average out while the retardation time also drops out (it just corresponds to a shift in the initial phase of the binary system) so that only remains the Newtonian limit discussed in~\ref{sec:newt}. This means that the binary-binary force is subdominant in this regime, as discussed in Section~\ref{sec:newt}.
     \item Binaries are close to a resonance $p n_1 = q n_2$ for some integers $p$ and $q$. Then the time-average can be nontrivial and the binary-binary force will probably induce order-one changes in the dissipative dynamics of the system if $\lambda \gg R$ or $\lambda \sim R$. We will discuss in more details the first case in Appendix~\ref{app}, while precisely evaluating the effect of the binary-binary force for $\lambda \sim R$ requires to solve a delay-differential equation, since in this generic setting the force involve retarded times. Delay-differential equations often present mathematical and numerical subtleties~\cite{10.1093/acprof:oso/9780198506546.001.0001}, so that we will leave the exploration of this regime to further work. 
 \end{itemize}

%\section{Numerical simulation?}\label{sec:num}

%\vphantom{test}

\section{Conclusions}

In this article we have derived the expression of a new kind of radiation-reaction force whose physical origin is related to an exchange of quadrupolar gravitons between two binary systems. Our expression generalizes the post-Newtonian formulae obtained in~\cite{jaranowskiRadiativePostNewtonianADM1997,koenigsdoerfferBinaryBlackholeDynamics2003,schaeferThreebodyHamiltonianGeneral1987} to generic cases when binaries are separated by distances equal or greater than the size of their Near Zone. However, to determine if the force can really be at play in the mechanism leading to the merging of binary systems requires solving a delay-differential equation which we leave to future work. It would be ultimately quite interesting to add the binary-binary force in $N$-body simulations of globular clusters or galactic nuclei such as~\cite{Rodriguez:2015oxa, Miller:2002pg, VanLandingham:2016ccd} and assess its impact on the merging rate of binary systems, or to re-examine the resonant capture in quadruple systems discussed in~\cite{2020MNRAS.493.5583T, Seto:2018xnu, 2018MNRAS.475.5215B, Seto:2020lez} taking into account the full expression of the binary-binary force. As we have argued, the new force probably necessitates the two binaries to be in resonance in order to give a non-negligible contribution on long timescales, which means that the modification to the $N$-body dynamics will only occur when one binary crosses the frequency of another one.
Another interesting possibility for phenomenological applications is the merger rate of primordial BH binaries: this rate has been shown to be quite sensitive to many-body effects~\cite{Jedamzik:2020ypm, Raidal:2018bbj} so that it would be interesting to know how these results could get changed taking into account the binary-binary force. Nevertheless, on the theoretical side the implications of the binary-binary force remain quite fascinating, as it shows that emission of gravitational waves from $N$-body systems presents some unique features which do not have any equivalent in the two-body problem.

%Our results could have several interesting applications to GW astronomy. 

\acknowledgments
%I would like to thank Guillaume Faye and Enrico Trincherini for comments on the draft. 
This research has been partly supported by the Italian MIUR under contract 2017FMJFMW (PRIN2017). I would like to thank an anonymous referee for pertinent remarks which led to the addition of Appendix~\ref{app}.

% \appendix
% \section{Formulaes}\label{sec:formulaes}

% \begin{align}\label{eq:Vijkl}
% \begin{split}
%     V_{ijkl}(k^0, \bm k) &= 4\pi G \big[ k_i k_j k_k k_l \\
%     &- \big(k^0\big)^2 \big(k_i k_k \delta_{jl} + k_i k_l \delta{jk} + k_j k_k \delta_{il} + k_j k_l \delta_{i k} \big) \\
%     &+ \big(k^0\big)^4 \big( \delta_{ik}\delta_{jl} + \delta_{il}\delta_{jk} \big) \big]
% \end{split}
% \end{align}
% where to derive this expression we have used the fact that the quadrupole moment is traceless to remove any term proportional to $\delta_{ij}$ or $\delta_{kl}$.

\appendix
\section{Averaged binary-binary force for $\lambda \gg R$}\label{app}

In order to give a simplified example of the averaging procedure discussed in Section~\ref{sec:long_timescales}, let us concentrate on the case $\lambda \gg R$ and derive an averaged radiation-reaction force. One immediately sees from Eq.~\eqref{eq:smallRLimit} that if the two binaries are not in a resonant state, then one can average separately each quadrupole moment over time and the effective action vanishes since $\langle \mathrm{d}^5 Q^{ij} / \mathrm{d}t^5 \rangle = 0$. On the other hand, when the two binaries are in a resonance we cannot split the total time-average in a product of averages over individual quadrupoles. Let us assume that the binaries are in a $1:1$ resonance. We denote by $a_1, e_1, \iota_1, \Omega_1, \omega_1, \eta_1$ the osculating elements of the first binary (resp. semimajor axis, eccentricity, inclination, longitude of ascending node, argument of periapsis, eccentric anomaly), and similarly for the second binary. For simplicity, we will assume circular trajectories $e_1=e_2=0$; furthermore, the $1:1$ resonance condition imposes $a_2 = a_1 (m_2/m_1)^{1/3}$ as well as $\eta_2 = \eta_1 + \phi$ where $\phi$ is a dephasing varying on secular timescales only. Finally, we will also assume $\Omega_2 = \Omega_1$ since we will get much simpler expressions in this case. Using the expression~\eqref{eq:smallRLimit} for the binary-binary effective action, one can use standard techniques as in e.g.~\cite{Kuntz:2021ohi} to derive the Lagrange planetary equations describing the time-evolution of all planetary elements. Averaging over one orbital timescale is then equivalent in our case to integrate the eccentric anomaly $\eta_1$ from $0$ to $2\pi$.
We are mostly interested in the averaged evolution equation for the semimajor axis, which can be used to define the merger timescale of the binaries. Including also the standard radiation-reaction terms for an isolated binary, it reads
\begin{align}
\begin{split} \label{eq:da1dt_general}
    \frac{\mathrm{d}a_1}{\mathrm{d}t} &= - \frac{64}{5} \frac{G^3 m_1^2 \mu_1}{a_1^3} \bigg[1+\frac{\mu_2 m_2^{2/3}}{\mu_1 m_1^{2/3}} \times \\
    &  \cos^4 \bigg(\frac{\iota_1 - \iota_2}{2} \bigg) \cos 2 \big( \phi+\omega_2-\omega_1 \big)  \bigg]
    \end{split}
\end{align}
where we recall that $m_1$ is the total mass of the first binary and $\mu_1$ its reduced mass. Of course, one can get the equation on $a_2$ just by exchanging $(1 \leftrightarrow 2)$. The binary-binary force manifests itself as the second term inside the square brackets, while the first one is just the standard Peter-Mathews formula.
We thus see that the merger time can a priori be greatly reduced with respect to the case of an isolated binary, especially if one has a hierarchy of mass $m_2 \gg m_1$. However, one should remember that a resonance condition has been assumed to derive Eq.~\eqref{eq:da1dt_general}; if any of the two semimajor axis changes substantially then the binaries will exit resonance and the radiation-reaction force will average to zero. %Thus, the ultimate effect of the binary-binary force on the merger time of binaries can be probably only studied through $N$-body simulations without averaging over the 

\bibliography{refs}% Produces the bibliography via BibTeX.

\end{document}